\documentclass[%
 reprint, 
nofootinbib,
 amsmath,amssymb,
 aps, physrev,
]{revtex4-2}

\usepackage{graphicx}% Include figure files
\usepackage{dcolumn}% Align table columns on decimal point
\usepackage{bm}% bold math
\usepackage{xcolor}
\usepackage{amsmath}
\usepackage{appendix}
\usepackage{chngcntr}
\usepackage[colorlinks=true, linkcolor=blue, citecolor=blue, urlcolor=blue]{hyperref}

\newcommand{\aap}{Astronomy \& Astrophysics}
\newcommand{\mnras}{Monthly Notices of the Royal Astronomical Society}
\newcommand{\apjl}{The Astrophysical Journal Letters}
\newcommand{\pasp}{Publications of the Astronomical Society of the Pacific}
\newcommand{\apjs}{The Astrophysical Journal Supplement Series}
\newcommand{\aj}{The Astronomical Journal}

\begin{document}
%TC:ignore

\title{Theoretical Predictions for the Inner Dark Matter Distribution in the Milky Way Informed by Simulations}

\author{Abdelaziz Hussein}
\affiliation{Department of Physics and Kavli Institute for Astrophysics and Space Research,
Massachusetts Institute of Technology, 77 Massachusetts Ave, Cambridge MA 02139, USA}

\author{Lina Necib}
\affiliation{Department of Physics and Kavli Institute for Astrophysics and Space Research, Massachusetts Institute of Technology, 77 Massachusetts Ave, Cambridge MA 02139, USA}
\affiliation{The NSF AI Institute for Artificial Intelligence and Fundamental Interactions}

\author{Manoj Kaplinghat}
\affiliation{Center for Cosmology, Department of Physics and Astronomy\\
University of California Irvine, CA 92697, USA}

\author{Stacy Y. Kim}
\affiliation{Carnegie Observatories, 813 Santa Barbara Street, Pasadena, CA 91101, USA}

\author{Andrew Wetzel}
\affiliation{Department of Physics \& Astronomy, University of California, Davis, CA 95616, USA}

\author{Justin I. Read}
\affiliation{Department of Physics, University of Surrey, Guildford, GU2 7XH, UK}
 
\author{Martin P. Rey}
\affiliation{Department of Physics, University of Bath, Claverton Road, Bath, BA27AY, UK}

\author{Oscar Agertz}
\affiliation{Lund Observatory, Division of Astrophysics, Department of Physics, Box 43, SE-221 00, Lund, Sweden}

\date{\today}% It is always \today, today,

\begin{abstract}

We build a theoretical range for the Milky Way's (MW) inner dark matter (DM) distribution informed by the FIRE-2, Auriga, {\small VINTERGATAN-GM}, and TNG50 simulation suites assuming the canonical cold dark matter (CDM) model. 
The DM density profiles in Auriga, {\small VINTERGATAN-GM}, and TNG50 can be approximately modeled using the adiabatic contraction prescription of Gnedin et al. 2004, while FIRE-2 has stronger baryonic feedback, leading to a departure from the adiabatic contraction model. 
The simulated halos that are adiabatically contracted are close to spherical (axis ratio $q \in [0.75-0.9]$ at $5^\circ$), whereas halos that experience strong baryonic feedback are oblate ($q \in [0.5-0.7]$). 
Using the adiabatic contraction and strong baryonic feedback models, along with the observed stellar distribution of the MW, the inner logarithmic density slope for CDM in the MW is predicted to range from $ -0.5$ to $-1.3$. 
The $J$-factor, which determines the DM-annihilation flux, averaged over a solid angle of $5^\circ$ ($10^\circ$) is predicted to span the range $0.8$--$30$ ($0.6$--$10$) $\times 10 ^{23} \rm{GeV}^2/\rm{cm}^5$. The $D$-factor, which determines the flux due to DM decay, is predicted to be in the range $0.6$--$2$ ($0.5-1$) $\times10^{23} \rm{GeV}/\rm{cm}^2$. 

GitHub: The results for this work can be found at \href{https://github.com/abdelazizhussein/MW-Inner-DM-Profile}{\includegraphics[height=1.2em]{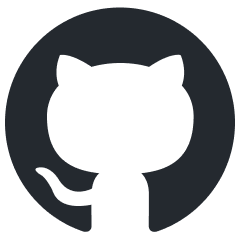}} \url{https://github.com/abdelazizhussein/MW-Inner-DM-Profile}.

\end{abstract}
\maketitle

\textit{Introduction}— Understanding the particle nature of dark matter (DM) is a central focus of modern physics, spanning both the areas of particle physics and astrophysics \cite{Mayet_2016,Gaskins_2016,Boveia_2018,Bertone-2018,Aschersleben_2024}. Precisely mapping its distribution within the inner few kiloparsecs of the Milky Way (MW) is critical for various detection strategies, especially indirect searches, that rely on the enhanced DM density near the Galactic Center (GC) to produce detectable Standard Model particles \cite{leane2020indirectdetectiondarkmatter,DM_review}.

Due to the extreme interstellar extinction \cite{2009ApJ...696.1407N,2010A&A...511A..18S,2011ApJ...737...73F}, source crowding towards the GC  \cite{
2018A&A...609A..26G}, and the fact that most of the mass in the GC is in stars \cite{Mroz_2019},  methods using stellar kinematics to infer the DM distribution have left the density within $\sim6 \ \rm{kpc}$ broadly unconstrained. Current predictions of the DM density within $6 \ \rm{kpc}$ of the GC from stellar kinematics are merely a consequence of the functional form for the analytical profile used to fit the DM distribution \citep{ou2023dark,Zhou_2023}.

In this \textit{Letter}, we employ a new approach to bracket the MW's inner DM profile using the observed stellar profile, along with  models calibrated to cosmological hydrodynamic simulations. 
Such simulations typically model DM by solving the collisionless Boltzmann equation coupled with Poisson's equation, while also modeling gas dynamics using the Euler equations \citep{Vogelsberger:2019ynw}. 
These hydrodynamic equations must be complemented by various astrophysical processes that shape the evolution of galaxies such as: gas cooling, star formation, and stellar feedback. Feedback is a term generally used to describe processes that return energy to the interstellar medium (ISM), such as radiation pressure, stellar winds, and supernovae (SNe) \citep{Agertz_2013}.

We begin by establishing a simplified framework to understand and model the mechanisms that shape the DM distribution. Generally, the DM distribution is determined by the depth of the gravitational potential created by baryonic matter (i.e. stars and gas), as well as star formation and feedback processes. In the absence of significant feedback from stars and/or Active Galactic Nuclei (AGN), the potential-well sourced by the baryonic matter results in the contraction of the DM halo. 
To describe this process, we utilize adiabatic contraction (AC).  AC is the response of a DM halo to the gravitational potential of accreted, on an infinitely small timescale (see Appendix \ref{app:AC_calculation}), baryons \cite{Blumenthal,sellwood_2014}. If we assume an initially spherically symmetric DM halo, then the angular momentum of individual orbits is conserved (see Appendix \ref{app:AC_calculation}), allowing us to solve for the final DM distribution given an initial DM and stellar distribution along with the final stellar distribution. This method enables us to utilize observations of the MW's stars to bracket its inner DM profile using models calibrated to simulations.

Introducing stellar and/or AGN feedback can significantly alter AC, even leading to an expansion of the inner DM halo. This will occur if the feedback drives significant, repeated, gravitational potential fluctuations on a timescale comparable to the local dynamical time \cite{Navarro_1996,Read_2005,Pontzen2012,Martizzi_2013}. However, modeling such feedback processes is challenging as they are implemented in simulations through effective subresolution (or subgrid) models \citep{Chaikin_2023}. Subgrid models are necessary due to the limited numerical resolution and high computational cost of simulations, as well as the fact that the physics of star formation remains an unresolved problem \citep{Naab_2017}. Moreover, subgrid models vary between simulation suites, resulting in key differences in the DM distribution (see Appendix \ref{sec:DM profiles}). More generally, evolving the same galaxy with different subgrid implementations can lead to variations in the stellar mass and gas distribution in the simulated galaxy at present day \citep{aquila_project,AGORA}. In spite of these variations due to subgrid and baryonic models, cosmological simulations have been successful at reproducing some key observational findings \cite{Millennium,Behroozi,Hopkins_2012,Moster,hopkins_2013,Narayanan_2013,Hopkins_2014,Hopkins_2014,SMHM,Kearn_2017,Kearn_2018,Kearn_2019,Auriga_1,Osinga_2024}. One must question, however, whether these successes are due to the implementation of the ``correct" model that captures all astrophysical process in nature, or if the errors offset each other's shortcomings enough to produce a seemingly correct prediction. This is a crucial point as placing constraints on the MW's inner DM profile requires a deep understanding of the intricate connection between DM and baryons \cite{Velmani_2023,velmani2024,Wu_2025}. In our effective picture, this corresponds to understanding if contraction or feedback-driven expansion is the dominant process that reshapes the DM distribution.

In this \textit{Letter}, we bracket the range of expected density profiles in the MW using 6 MW mass galaxies from the Auriga L3 \citep{Auriga_1}, {\small VINTERGATAN-GM} \citep{vintergatan, Vintergatan_GM}, TNG50 \citep{TNG_50}, and FIRE-2 (Latte suite) \citep{FIRE_2} simulations. We achieve this by characterizing the 
 effect of the different implementations of feedback, in the various simulation suites, on the DM profile relative to AC. We show that the DM density profiles observed in Auriga, {\small VINTERGATAN-GM}, and TNG50 can be well modeled using the AC model of Gnedin et al. 2004 \cite{Gnedin_2004}, hereafter G04, while FIRE-2 has a stronger feedback effect that leads to an expansion of the inner DM halo (see \cite{Lazar_2020} for an analytical fit of the DM profiles in the FIRE-2 simulations). 
We further characterize the morphology (i.e., deviation from spherical symmetry) of the DM density profiles and find that spherical symmetry is a good approximation for the adiabatically-contracted profiles, while FIRE-2 predicts an ellipsoidal shape aligned with the disk.

\begin{table*}[t]
    \centering
    \begin{tabular}{ccccccccccc}
        \hline
        Simulation(s) & Code & Technique & SMBH & ${M}_{\mathrm{DM}}$ & ${M}_{\mathrm{baryon}}$ & $\epsilon_{\mathrm{DM}}$ & $\epsilon_{\mathrm{stars}}$ & Reference(s) \\
         &  &  & Feedback & $[{M}_\odot]$ & $[{M}_\odot]$ &  [pc] & [pc] &  \\
        \hline
        Auriga L3 & AREPO & zoom & yes  & $5 \times 10^4$ & $6 \times 10^3$ & 188  & 188 & \citep{Auriga_2} \\
        FIRE-2 Suite & GIZMO & zoom & no  & $3.5\times10^4$  & $7.1\times10^3$ & 40 & 4 & \citep{FIRE_2} \\
        {\small VINTERGATAN-GM} & RAMSES & zoom & no  & $2\times10^5$ & $3\times10^4$ & 20 & 20 & \citep{Vintergatan_GM} \\
        TNG50 & AREPO & unif.res. & yes & $4.5\times10^5$ & $8\times10^4$ & 288 & 288 & \citep{TNG50_01} \\
        \hline
    \end{tabular}
    \caption{The simulations used in this paper and their parameters are listed above. The parameters are as follows from left to right: the code used to solve the hydrodynamic equations, the simulation type (zoom-in or uniform resolution box), the implementation of SMBH Feedback, ${M}_{\mathrm{DM}}$ and ${M}_{\mathrm{baryon}}$ are the DM and baryon particle mass respectively, and $\epsilon_{\mathrm{DM}}$ along with $\epsilon_{\mathrm{stars}}$ are the DM and stellar softening lengths in kpc respectively.}
    \label{tab:sim_info}
\end{table*}

We summarize some key features of each simulation suite in Table \ref{tab:sim_info} (we defer a detailed discussion of the methods used to solve the hydrodynamical equations, parameters that characterize the simulation run, as well as a description of the astrophysical processes modeled to Appendix \ref{sec:simulations}). We further outline details about the 6 MW analogs chosen from each simulation suite that will be used throughout the analysis in Table \ref{tab:consolidated}.

\begin{figure}[b]
    \centering
    \includegraphics[width=0.5\textwidth]{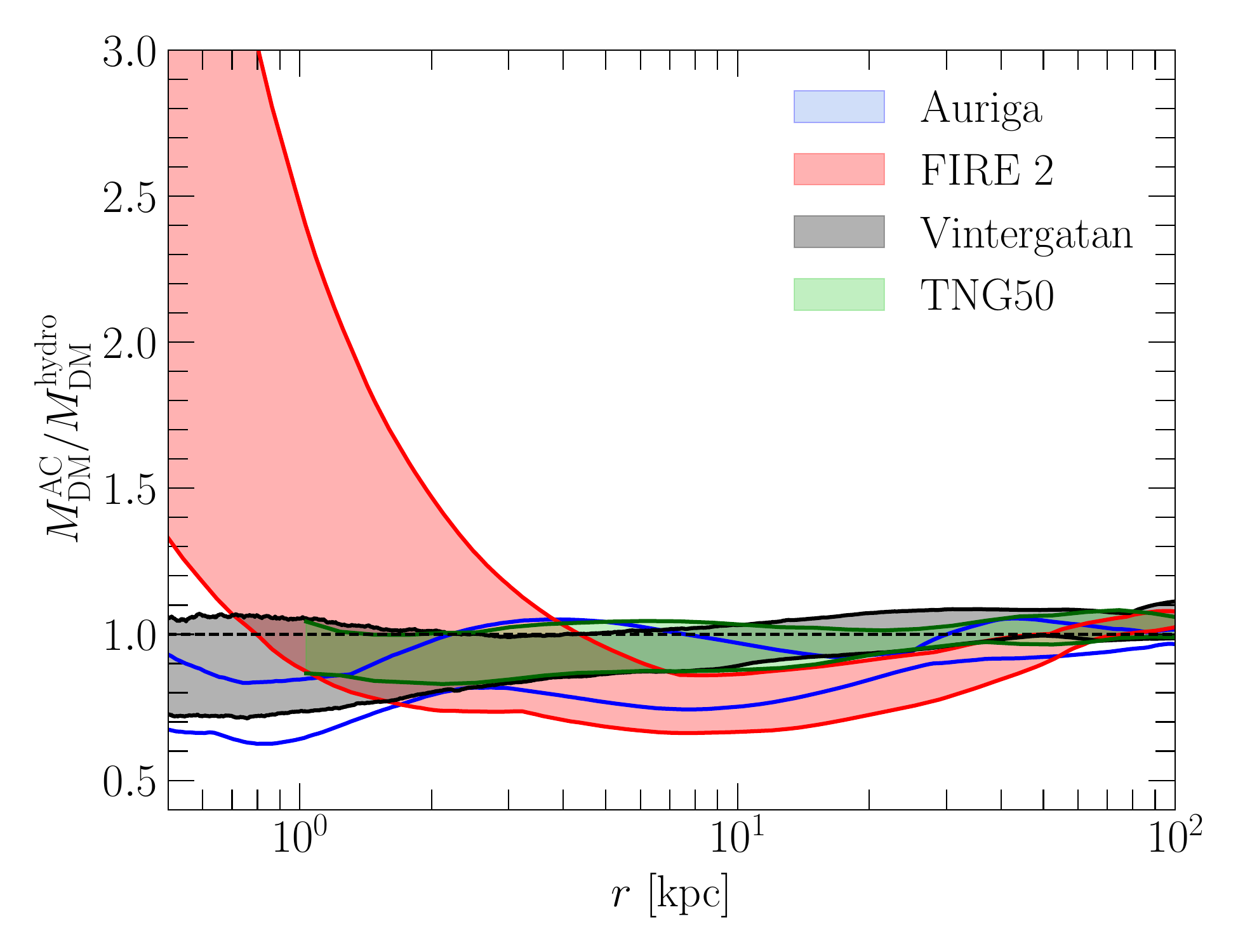}
    \caption{Ratio of ${M^{\text{AC}}_{\text{DM}}}/{M^{\text{hydro}}_{\text{DM}}}$ for the various simulation suites. The band encompasses the halo to halo scatter. The calculation for TNG50 is halted at $1 \ \rm{kpc}$ due to the limited mass and spatial resolutions.}
    \label{fig:AC_hydro_ratio}
\end{figure}

\textit{Adiabatic Contraction}—To set bounds on the predicted DM profile for the MW, we begin by characterizing the feedback models relative to AC. We compare the AC prediction (outlined in Appendix \ref{sec:AC calc}) for the DM enclosed mass $M^{\text{AC}}_{\text{DM}}$ to that of the hydrodynamic simulations $M^{\text{hydro}}_{\text{DM}}$ by taking the ratio of the two ${M^{\text{AC}}_{\text{DM}}}/{M^{\text{hydro}}_{\text{DM}}}$ as shown in Figure \ref{fig:AC_hydro_ratio}. For $r\gtrsim 10~\text{kpc}$, ${M^{\text{AC}}_{\text{DM}}}/{M^{\text{hydro}}_{\text{DM}}} \approx 1$ with some scatter, consistent with the AC  model. As we probe smaller radii, we observe two solutions. For Auriga, {\small VINTERGATAN-GM} and TNG50, the ratio is $0.7<{M^{\text{AC}}_{\text{DM}}}/{M^{\text{hydro}}_{\text{DM}}} < 1.1$, highlighting that contraction is driving the dynamics. However for FIRE-2, the ratio is $0.7<{M^{\text{AC}}_{\text{DM}}}/{M^{\text{hydro}}_{\text{DM}}} < 3$ for $r\lesssim 10 \ \text{kpc}$. ${M^{\text{AC}}_{\text{DM}}}/{M^{\text{hydro}}_{\text{DM}}} >1$ indicates that AC is over predicting the DM mass at the given radius. We therefore find that the modeling of baryonic physics in FIRE-2 leads to a break away from the AC prescription. A large component of this break away can be attributed to stellar feedback from SNe and stellar winds, which alter the DM distribution. We describe the solution represented by FIRE-2 as strong baryonic feedback (SBF). 

\begin{figure*}[t]
    \centering
    \includegraphics[width=1\linewidth]{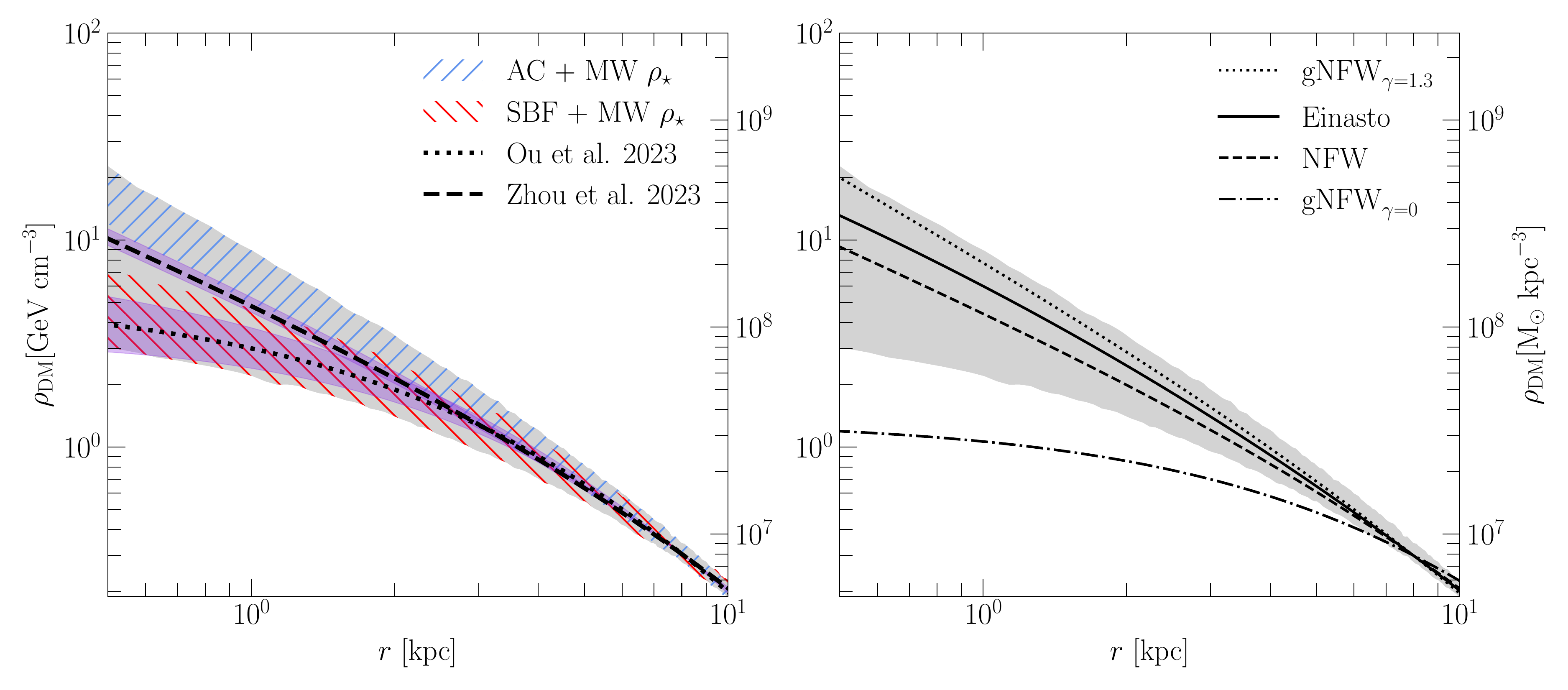}
    \caption{On the left, the prediction of the MW DM density profile with SBF in red and AC in blue. We bracket these two extrema with the band in gray. We over plot two fits of the MW's $\rho_\mathrm{DM}$ resulting from circular velocity curve measurements (see \cite{ou2023dark} and \cite{Zhou_2023} more details). On the right, we plot the same bracketed range along with four suggested $\rho_\mathrm{DM}$ from \cite{Reinert_2018} and \cite{winos} that are used frequently in particle physics analyses. Note that the $\rho_{\rm{gNFW}}(\rho_0,r_s,R_0,\gamma)$ profiles have slightly different values for $\rho_0,r_s$ in addition to the different values of $\gamma$ listed (see Table 5 of \cite{Reinert_2018}). Three of the four profiles fall within the proposed range, while the cored gNFW profile with $\gamma = 0$ is well outside this range}
    
    \label{fig:MW_prediction}
\end{figure*}

\textit{Bounds on the predicted DM profile ($\rho_{\rm{DM}}$) of the MW }—We now focus our attention to setting bounds informed by simulations on the MW's $\rho_{\rm{DM}}$. To do so, we use AC and SBF as the two limiting cases for the possible dominant mechanisms for redistributing DM in the inner Galaxy. To predict the MW's $\rho_{\rm{DM}}$ for the AC case, we use the initial $\rho_{\rm{DM}}$ from the dark matter only (DMO) FIRE-2 simulations\footnote{The choice is generic as the DMO profiles between the various simulation suite are generic NFWs and do not vary substantially in shape.} along with the MW's observed stellar distribution as input (see Appendix \ref{sec:AC calc}). Specifically, the analytical fit for the observed stellar distribution from \cite{Model_2}, where the density profile is modeled using a bulge, along with thin and thick disks. 
To obtain the SBF prediction, we multiply the result from AC by the corresponding SBF ratio of ${M^{\text{hydro}}_{\text{DM}}}/{M^{\text{AC}}_{\text{DM}}}$. The results are shown in Figure \ref{fig:MW_prediction}, where we additionally plot the inferred MW DM density from \cite{ou2023dark} and \cite{Zhou_2023}, which utilize the excellent stellar kinematic data from the latest \textit{Gaia} mission data release \cite{2016A&A...595A...1G,Gaia_DR3}. The hatchings in the figure show the halo to halo variations, while the gray band encompasses the space of possible $\rho_{\rm{DM}}$ for the MW. We note that our $\rho_{\rm{DM}}$ is not defined below $0.2 \ \rm{kpc}$, or $ 2^\circ$ from the GC, due to the simulations' resolution and to ensure a convergent result from our AC calculation. The band is available for public use at 
 \href{https://github.com/abdelazizhussein/MW-Inner-DM-Profile}{\includegraphics[height=1.2em]{logos/github-mark.png}} \url{https://github.com/abdelazizhussein/MW-Inner-DM-Profile} .

The band encompasses inner logarithmic slopes ranging from $-0.5$ to $-1.3$.\footnote{The inner logarithmic slopes described here are those of a generalized NFW Profile \cite{gNFW}. These values of $\gamma$ are for a $\rho_{\rm{gNFW}}(\rho_0,r_s,R_0,\gamma)$ with $\rho_0 = 0.45 \ \rm{GeV}/\rm{cm}^3$, $ r_s = 17.2 \ \rm{kpc}$, $R_0 = 8.2\ \rm{kpc}$ as defined in \cite{Reinert_2018}.} We find that the fitted $\rho_{\rm{DM}}$ for \cite{ou2023dark} and \cite{Zhou_2023} lie within the predicted band for the simulations. However, we caution that within $ \sim 6 \  \rm{kpc}$ the ``observed" profiles from \cite{ou2023dark} and \cite{Zhou_2023} are an extrapolation dictated by the analytical model used to fit $\rho_{\rm{DM}}$ due to the difficulty in measuring the kinematics of the bulge to infer $\rho_{\rm{DM}}$ (see \cite{ou2023dark} for more details). We further plot the DM profiles utilized frequently in particle physics analyses \cite{Reinert_2018,winos}. While these analyses use a wide range of profiles, some may predict DM signals outside the range expected from the competing effects of AC and SBF, making them less likely.

We emphasize that our novel prediction range is informed by simulations, and steps away from simply utilizing $\rho_{\rm{DM}}$ extracted from simulations as has been previously done in the literature \cite{Calore_2015, Bernal_2016, rodd2024} or using superficially motivated ``cuspy" or ``cored" profiles \cite{Modak_2015,Blanger_2022, boto2024}. Rather, we bracket the MW's $\rho_{\rm{DM}}$ using the two limiting cases identified in simulations that shape $\rho_{\rm{DM}}$: AC and SBF. We advise using multiple density profiles that fall within the range outlined in Figure \ref{fig:MW_prediction} for analyses.

\begin{figure*}[t]
    \centering
    \includegraphics[width=1\linewidth]{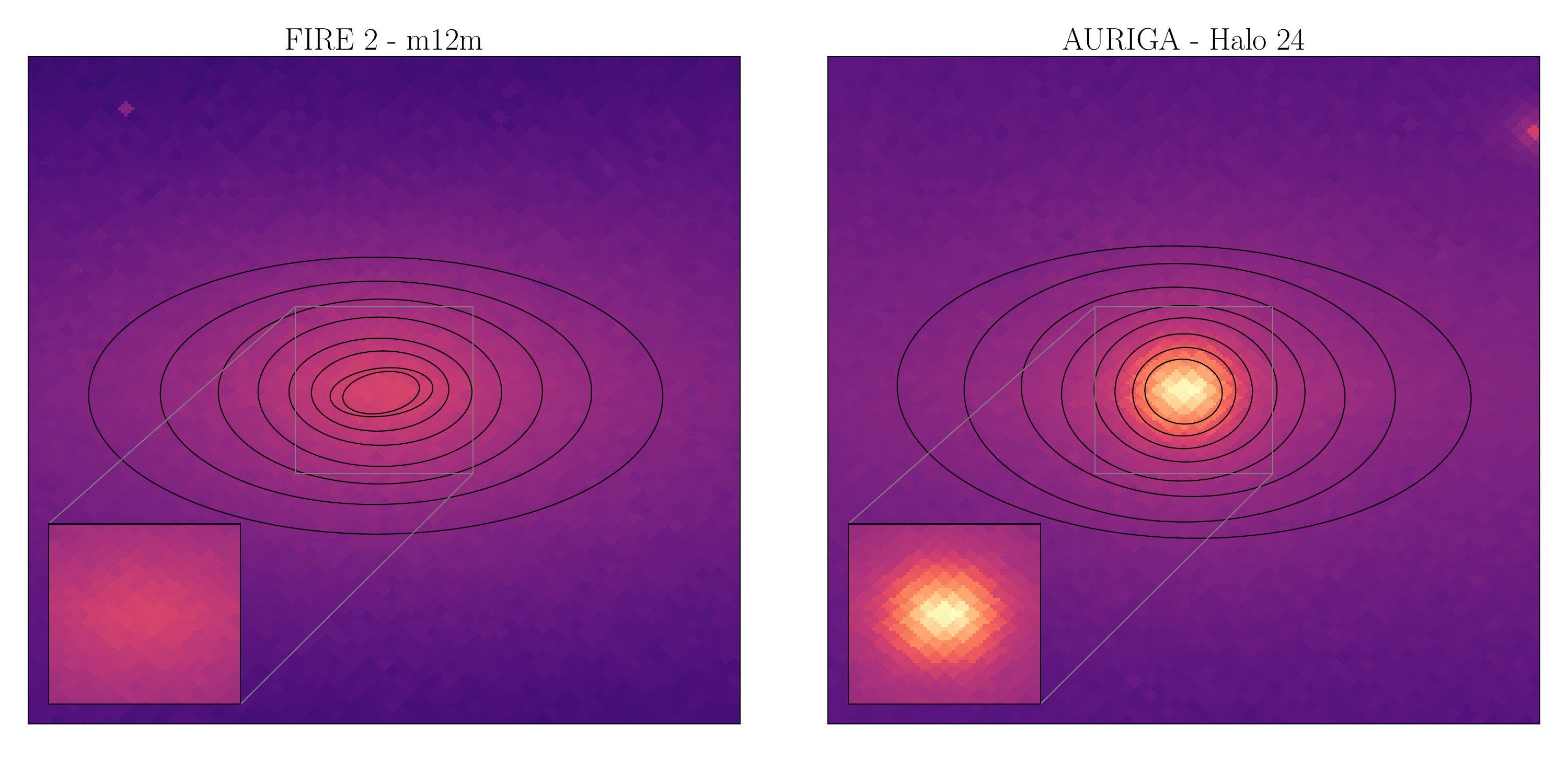}
    \caption{\label{fig:Healpix_plot}$40\times 40^\circ$ annihilation flux map normalized to the same scale with isophote contours. On the left is FIRE-2 m12m, and on the right is Auriga Halo 24. An inset for the $10 \times 10^\circ$ field of view is inserted at the bottom left corner of each plot.}
\end{figure*}

\textit{$J$ and $D$ factors calculation}—To get a sense of how the limiting cases of AC and SBF affect DM searches, we calculate the potential DM indirect detection signal based on the bracketed range of density profiles. Generically, for models where DM self-annihilates (or decays) to produce a signal, the annihilation flux is $\propto \rho^2_{\rm{DM}}$ ($\propto \rho_{\rm{DM}}$).

We calculate the annihilation flux or the $J$ factor through the following line of sight ($l.o.s$) integral (see Appendix \ref{sec: flux map calculation} for further details):
\begin{equation} \label{eq:los_integral}
    J = \int_{l.o.s} \bigg(\frac{\rho}{\rho_\odot}\bigg) ^2 \frac{ds}{r_\odot}
\end{equation}
and similarly the decay flux or the $D$ factor:
\begin{equation} \label{eq:D_factor_integral}
    D = \int_{l.o.s} \bigg(\frac{\rho}{\rho_\odot}\bigg) \frac{ds}{r_\odot}
\end{equation}
 where $r_\odot = 8 \ \rm{kpc} $, and $\rho_\odot= 0.3 \ \rm{GeV/cm^3}$ \cite{Marco_Cirelli_2011}. 
  We calculate the $J$ and $D$ factors averaged over solid angle of $5^\circ(10^\circ)$ using the bounds of our bracketed range in Figure \ref{fig:MW_prediction}.  We adopt a conservative estimate on the $J$ and $D$ factors by asserting that $\rho_{\rm{DM}}(<0.2\ \rm{kpc}) = \rho_{\rm{DM}}(0.2 \ \rm{kpc})$. We find that the annihilation signal predicted by the upper bound of our bracketed density range is two (one) orders of magnitude stronger than that predicted by our lower bound DM density at 
  $3\times 10 ^{24}$ and $8\times 10 ^{22}$ ($1\times 10 ^{24}$ and $6\times 10 ^{22}$)  $\rm{GeV}^2/\rm{cm}^5$ respectively. We further find the decay flux to span the range $0.6$--$2$ ($0.5-1$) $\times10^{23} \rm{GeV}/\rm{cm}^2$.

 \textit{Annihilation flux map}— We further explore the shapes of the DM halos in the simulations by constructing annihilation flux maps.
 These maps are constructed by integrating Equation \ref{eq:los_integral} along lines of sight within a $40\times 40^\circ$ region around the GC (see Appendix \ref{sec: flux map calculation} for details). The results are shown in Figure \ref{fig:Healpix_plot} for the FIRE-2 latte suite (representing SBF) and Auriga L3 suite (representing AC). We observe that the brightest pixel in the AC case (right panel) is $\approx 20\times$ brighter than the SBF case (left panel). 

Using the annihilation flux maps, we fit ellipses to contours of constant flux to infer the shape of the inner DM halo (see Appendix \ref{sec: axis ratios} for details). We excluded TNG50 from this analysis due to its limited spatial resolution, and {\small VINTERGATAN-GM} due to the presence of numerous subhalos within the inner galaxy which complicate the process of fitting ellipses (see Appendix \ref{sec:Vintergatan_flux}). The constant flux contours at $5^\circ$ from the GC tend to be spherical for Auriga (axis ratio q in the range $0.75-0.9$), while they exhibit an oblate morphology for FIRE ($\rm{q}: 0.5-0.7$). Outside of $10^\circ$, the FIRE halos tend to have a constant q which is more spherical than the inner $10^\circ$, while the Auriga halos become more ellipsoidal (see Figure \ref{fig:axis-ratios}).

Fits to the Galactic Center Excess (GCE) (see \cite{annurev_GCE} for details), if interpreted as DM annihilation flux, generally prefer steep density profiles (log-slopes $\sim-1.2$) and approximate spherical symmetry \cite{Murgia:2020dzu,Di_Mauro_2021}. The lack of spherical symmetry and shallow DM density profiles in the FIRE halos suggest that the spatial template of a potential signal from DM annihilation can be significantly more varied, as illustrated in Figures \ref{fig:MW_prediction} and \ref{fig:Healpix_plot}. Although elliptical shapes and shallower density profiles have been explored in the literature \cite[see e.g.][]{Daylan_2016,Abazajian:2020tww}, their ranges and correlations have not been elucidated previously based on simulations. 

To understand the combination of density slopes and shapes to be used as templates, we explored the correlation between the two variables in the simulations. There is a weak correlation between $\rm{q(5^\circ)}$ and the average density of the DM within 1 kpc, and a moderate anti-correlation with the stellar density at 1 kpc (see Figure \ref{fig:rho_q_correlation}). 
Our results suggest that a spherically symmetric distribution should be accompanied with a steeper inner slope, as has often been assumed in the literature, whereas an oblate morphology should be paired with a shallower inner profile. 
We emphasize that the flattening of the spherical halo into an ellipsoid is not solely due to AC but also stellar processes, for example, angular momentum transport due to the formation of the stellar disk and galactic bar~\cite{Petersen_2016}.

\textit{Summary}—In this \textit{Letter}, we studied how different simulation suites compare with the physically motivated adiabatic contraction (AC) model of G04. We found that AURIGA, {\small VINTERGATAN-GM}, and TNG50 closely match the expected results obtained from AC. FIRE-2, however, breaks away from AC largely due the different implementation of baryonic process (for example feedback from stellar winds, and supernovae (SNe)) leading to the SBF solution.

Using the two limiting cases of AC and SBF, along with observations of the MW's stellar distribution, we bracket the MW's $\rho_{\rm{DM}}$ (shown in Figure \ref{fig:MW_prediction}). We find that $\rho_{\rm{DM}}$ has an inner slope roughly between $-0.5$ and $-1.3$.
The range encompasses measurements of the MW's $\rho_{\rm{DM}}$ inferred from circular velocity measurement \cite{ou2023dark, Zhou_2023}, though the results within $\sim6 \ \rm{kpc}$ are extrapolated towards the center of the Galaxy.

To illustrate how the limiting cases impact potential DM signals, we focus on the case of indirect detection searches. We calculate the $J$ factor for the limiting cases of AC and SBF to find values averaged over a solid angle with an opening of $5^\circ$($10^\circ$) to be in the range  $0.8$--$30$ ($0.6$--$10$) $\times 10 ^{23} \rm{GeV}^2/\rm{cm}^5$, and the $D$ factor to span the range $0.6$--$2$ ($0.5-1$) $\times10^{23} \rm{GeV}/\rm{cm}^2$ .

We further analyzed the shape (i.e. deviation from spherical symmetry) of the DM halos in the simulations and found that spherical symmetry is a good approximation for the adiabatically-contracted profiles with axis ratios ranging from $0.75-0.9$, while FIRE-2 predicts an oblate shape aligned with the disk with axis ratios ranging from $0.5-0.7$.

A spherically symmetric DM halo with an inner slope of $-1.2$ is often assumed in the literature; however, our results show that less steep profiles with oblate morphologies (and steeper profiles with spherical shapes) should also be explored as viable DM halo templates. There is a weak correlation between 
q$(5^\circ)$ and $\rho_{\rm{DM}}$ as well as an anti-correlation with $\rho_{\rm{stellar}}$ (see Figure \ref{fig:rho_q_correlation}) which can inform analysis that are highly sensitive to the halo shape (for example; weak lensing).

For calculations requiring the density profile, such as calculating the DM flux from the GC, we recommend a conservative systematic uncertainty bracketed by the prediction range in Figure \ref{fig:MW_prediction}.

As the field of cosmological simulations moves forward, we expect further advancements in refining the physics implementations of stellar feedback, inching us closer to a better understanding of galaxy formation. These numerical developments will also be met with efforts from observations to constrain the profile within $\sim 6$ kpc of the galactic center by peering through the dust to study the kinematics of stars in the bulge using missions such as the James Web Space Telescope (JWST). Therefore, we expect that the predicted range of DM density profiles will narrow as we gain more insight about the center of our Galaxy.

%TC:ignore
\section*{Acknowledgments}
We would like to thank Robert Grand, 
Dawei Zhong, Cian Roche, Rafael Boto, Viraj Pandya, Nicholas L. Rodd, Nathaniel Starkman, Maya Silverman, Mariangela Lisanti, Sandip Roy, Adriana Dropulic for helpful conversations.

This project was supported by the NSF award 2307788.
LN is supported by the Sloan Fellowship, the NSF CAREER award 2337864, NSF award 2307788, and by the NSF award PHY2019786 (The NSF AI Institute for Artificial Intelligence and Fundamental Interactions, http://iaifi.org/). MK is supported by the U.S. National Science Foundation (NSF) grant PHY-2210283. JIR would like to acknowledge support from STFC grants ST/Y002865/1 and ST/Y002857/1. AW received support from NSF, via CAREER award AST-2045928 and grant AST-2107772, and HST grant GO-16273 from STScI.
 OA acknowledges support from the Knut and Alice Wallenberg Foundation, the Swedish Research Council (grant 2019-04659), and the Swedish National Space Agency (SNSA Dnr 2023-00164).


\vspace{5mm}


{This research is part of the Frontera computing project at the Texas Advanced Computing Center. Frontera is made possible by National Science Foundation award OAC-1818253. The authors acknowledge the Texas Advanced Computing Center (TACC) at The University of Texas at Austin for providing computational resources that have contributed to the research results reported within this paper \url{http://www.tacc.utexas.edu}.
This research is part of the Frontera computing project at the Texas Advanced Computing Center. Frontera is made possible by National Science Foundation award OAC-1818253. 

We use simulations from the FIRE-2 public data release \cite{FIRE_2}. The FIRE-2 cosmological zoom-in simulations of galaxy formation are part of the Feedback In Realistic Environments (FIRE) project, generated using the Gizmo code \cite{Hopkins_2015b} and the FIRE-2 physics model \cite{Hopkins_2018}. We have used simulations from the Auriga Project public data release \citep{Auriga_1} available at \url{https://wwwmpa.mpa-garching.mpg.de/auriga/data}. We used simulations from the TNG50 public data release \citep{Pillepich_2019,Nelson_2019,TNG_50_public_release_2021} and Milky Way+Andromeda (MW/M31) Sample \url{https://www.tng-project.org/data/milkyway+andromeda/} \cite{pillepich2023milky}. The IllustrisTNG simulations were undertaken with compute time awarded by the Gauss Centre for Supercomputing (GCS) under GCS Large-Scale Projects GCS-ILLU and GCS-DWAR on the GCS share of the supercomputer Hazel Hen at the High Performance Computing Center Stuttgart (HLRS), as well as on the machines of the Max Planck Computing and Data Facility (MPCDF) in Garching, Germany.

We have also derived results using the healpy and \texttt{HEALPix} package \cite{healpix1,healpix2}. This research made use of \texttt{PHOTUTILS}, an Astropy \cite{astropy:2013,astropy:2018,astropy:2022} package for detection and photometry of astronomical sources \citep{photutils}.}

\bibliography{apssamp}% Produces the bibliography via BibTeX.

\clearpage
\newpage

\appendix

\counterwithin{figure}{section}
% \makeatletter

%\renewcommand{\thetable}{EM\arabic{table}}

\section{Bluementhal's Adiabatic Contraction}\label{app:AC_calculation}

The Lagrangian per unit mass for a particle in a circular orbit is 
\begin{equation}
    \mathcal{L} = \frac{1}{2} r^2 \left(\frac{d\theta}{dt}\right)^2 + \frac{G M}{r},
\end{equation}
where $G$ is the gravitational constant, $M$ is the mass of the central object the particle is orbiting around, $r$ is the distance from the center of the circular orbit and $\theta$ is the azimuthal angle.
Using the Euler Lagrange equations, we can show that the angular momentum $L$ is conserved:
\begin{equation}
    \frac{\partial\mathcal{L}}{\partial\theta} = \frac{d}{dt}\frac{\partial \mathcal{L}}{\partial\dot{\theta}} \rightarrow \frac{d}{dt}(r^2 \dot\theta) = 0 \rightarrow L = r^2 \dot\theta = \rm{const}
\end{equation}

If we assume that the potential ($V$) varies as the particle moves closer to the central mass over a time interval $t_i \leq t \leq t_f $ where the evolution is ``slow," in the sense that the potential's change is very small over a single orbital period $T_{\rm{orb}}$ \citep{scott}:
\begin{equation}
    \Bigg\lvert \frac{T_{\rm{orb}}}{V}\frac{\partial V}{\partial t}\Bigg\rvert \ll 1
\end{equation}

It can be shown (see appendix A of \cite{scott}), that the following integral of motion remains nearly constant when the hamiltonian changes adiabatically. Namely
\begin{equation}
    J_k = \frac{1}{2 \pi} \oint p_k dx^k,
\end{equation}
where $k$ labels the canonical coordinates used in the problem. For our purposes, we are interested in the angular action $J_\theta = \oint p_\theta d\theta$ where $p_\theta = L$ in this case. Thus 
\begin{equation}
    L = r^2 \dot \theta = r v_c = m r \sqrt{\frac{G M}{r^2}} = \sqrt{G M r}. 
\end{equation}
Therefore, we find that
\begin{equation}
M(r) r = \rm{constant}.
\end{equation}

\section{Dark Matter Density Profiles} \label{sec:DM profiles}
\begin{figure}[h]
    \centering
    \includegraphics[width=0.45\textwidth]{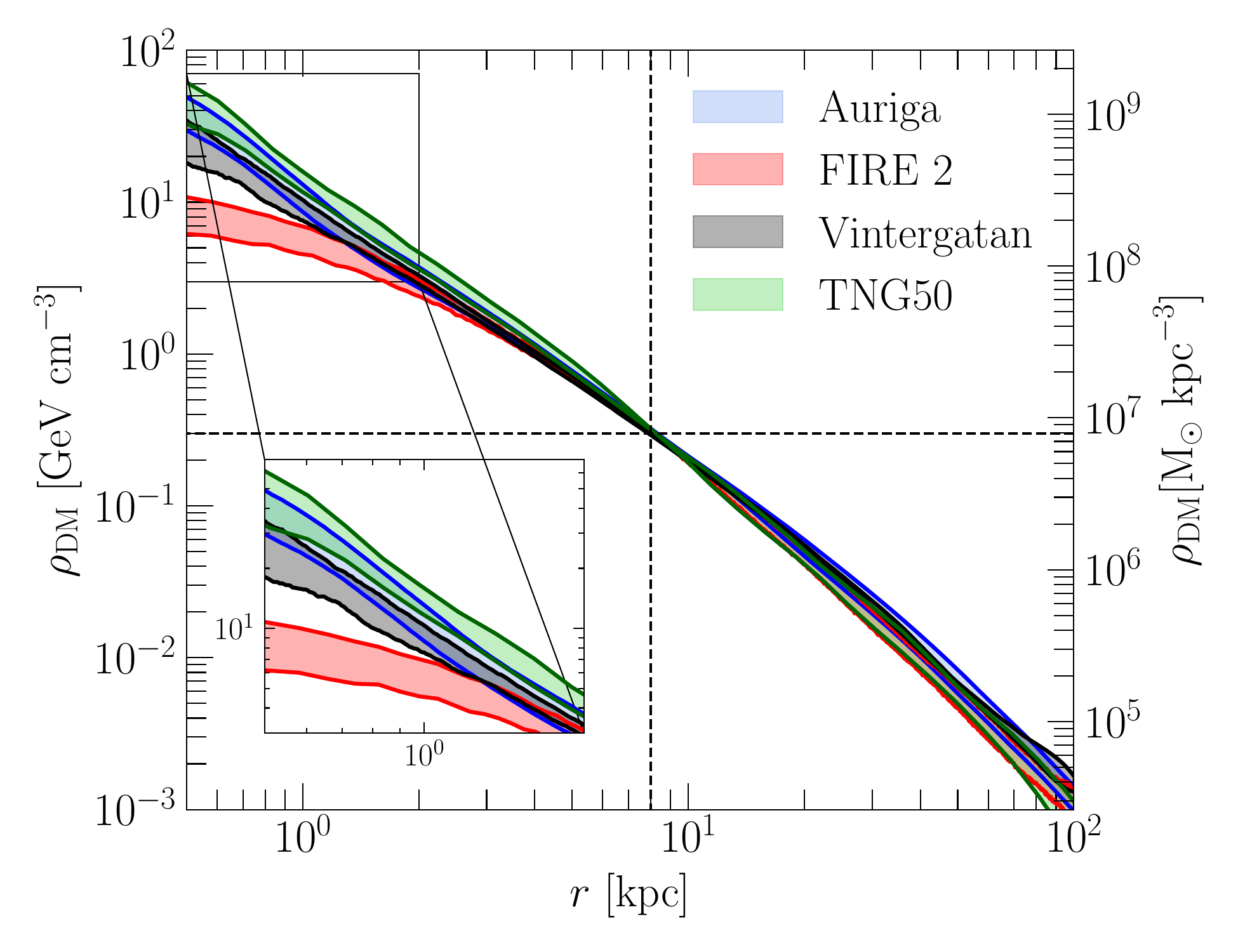}
    \caption{ 
    DM density profiles for the various simulation suites. For each suite, we encompass the DM density profile for all 6 halos in a band. The density profiles are  normalized to $0.3 \ \mathrm{GeV \ \mathrm{cm}^{-3}}$ at $8 \ \mathrm{kpc}$ indicated by the dashed lines.}
    \label{fig:dm_density}
\end{figure}

 We  examine the density profiles for each simulation suite in Figure \ref{fig:dm_density}. Each profile is normalized to $0.3 \ \mathrm{GeV \ \mathrm{cm}^{-3}}$ at $8 \ \mathrm{kpc}$ to match the local DM density in the solar neighborhood \citep{Eilers_2019}. The profiles are plotted as bands, where each band encloses the profiles for the 6 individual halos. We note that the density profiles from the different simulation suites are consistent with each other for $r \gtrsim4 \ \mathrm{kpc}$. However, for $r \lesssim 4 \ \mathrm{kpc}$, we identify different behavior exhibited by each simulation suite.
 We see that FIRE exhibits a ``cored" profile with an inner slope that is less steep than the ``cuspy" profiles of Auriga, TNG50 and {\small VINTERGATAN-GM}. 

\section{Simulations} \label{sec:simulations}
 We introduce each simulation suite in the following subsections by discussing the method used to solve the hydrodynamical equations, the size of the box that was simulated as well as the cosmological parameters used, followed by a brief description of the astrophysical processes modeled and some details pertaining to each of the simulations. Additionally, we include details about six MW analogs chosen from each simulation suite that will be used throughout the analysis (Table \ref{tab:consolidated}).

\subsection{FIRE-2 Simulations}

The FIRE-2 simulations is a set of cosmological zoom-in simulations performed with the multi-method gravity plus hydrodynamics code GIZMO \citep{Hopkins_2015b} \footnote{Though an updated stellar evolution and sub grid-model has been released by the collaboration in the form of FIRE-3 \citep{FIRE-3}, we are faced with two drawbacks that would hinder our current analysis: Only 3 different halos are currently available (compared to more than double that with FIRE-2), and the results of the new feedback model are still being investigated}. We specifically use the \textit{Latte} suite \citep{latte}, where the halos were picked from a $85.5 \  \mathrm{Mpc}$ width box with $\Lambda$CDM cosmology: $\Omega_m = 0.272$, $\Omega_b = 0.0455$, $\Omega_\Lambda = 0.728$, $h = 0.702$ \citep{Planck_2016}.

This suite utilizes the feedback implementation described in \cite{Hopkins_2018} and the mesh-free Lagrangian Godunov (MFM) method. The MFM approach provides adaptive spatial resolution and maintains conservation of mass, energy, and momentum. FIRE-2 includes radiative heating and cooling for gas across a temperature range of $10 - 1000 
\ \mathrm{K}$. Heating sources include an ionising background \citep{Faucher_2009}, stellar feedback from OB stars, AGB mass-loss, type Ia and type II supernovae, photoelectric heating, and radiation pressure, with inputs taken directly from stellar evolution models. The simulations self-consistently generate and track 11 elemental abundances (H, He, C, N, O, Ne, Mg, Si, S, Ca, and Fe), and include subgrid diffusion of these elements in gas via turbulence \citep{Hopkins_2015,Su_2017,Escala_2017}. Star formation occurs in gas that is locally molecular, self-gravitating, sufficiently dense ($>$ 1000 $\mathrm{\mathrm{cm}}^3$), and Jeans unstable  (following \cite{Krumholz_2011}).

Stellar feedback is modeled via local deposition of mass, momentum, energy, and
metal mass from star particles to neighboring gas particles. The
feedback accounts for both Type Ia and Type II SNe, stellar
winds, momentum from radiation pressure, photoionization, and
photoelectric heating. In this way, the generation, propagation,
and recycling of large-scale galactic winds are emergent
phenomena rather than being put in “by hand” via delayed
cooling, thermal bombs, or decoupled winds \citep[e.g.][]{Muratov_2015,Angles_2017}.

\subsection{{\small VINTERGATAN-GM} Simulations}

We use the \small{VINTERGATAN-GM} suite \citep{Vintergatan_GM} of simulations, performed with the adaptive mesh refinement code RAMSES \citep{Teyssier_2002} assuming a flat CDM cosmology with $h =0.6727$, $\Omega_m = 0.3139$, $\Omega_b = 0.04916$ \citep{Planck_2016} and linearly span cosmic
time between $z = 99$ and $z = 0$. The fiducial simulation of the suite (685) focuses on a MW-mass system, whereas the
other four simulations (685x09, 685x095, 685x11, 685x12) are variations where the initial conditions were genetically
modified (GM) to alter the mass ratio of an important merger that
occurs at $z \approx 2$, while maintaining the $z = 0 $ halo mass of the system \citep[see][]{Vintergatan_GM},
through the use of the GENETIC code \citep{Roth_2015,
Rey_2017,Stopyra_2021}. We emphasize that we are using the GM suite rather than the original simulation run presented in \citep{vintergatan}. The original simulation was run at a higher  resolution and features a more massive disk compared to the GM suite. Other simulations, not included in this analyses, in the GM suite do feature more massive discs, e.g. simulation 599. It is important to note that simulation 715, included in this analysis, yields an early type galaxy.

The galaxy formation model \citep[used in the GM suite and outlined in ][]{vintergatan} includes prescriptions for star formation, feedback from SNe Type Ia and
SNe Type II, and stellar winds from O, B, and AGB stars. Stars are formed
from cold dense gas ($\rho > 100 \mathrm{cm}^{-3}, T < 100 \mathrm{K}$) generating stellar
particles with an initial mass of $10^4 \mathrm{\mathrm{M}_\odot}$, modeled with a \cite{Chabrier_2003} initial mass function. Feedback is injected in the form of
thermal energy when the cooling radius is resolved by at least six
gas cells, and in the form of momentum otherwise \citep{kim_2015,Martizzi_2015,vintergatan}. 

\subsection{TNG50 Simulations}\label{subsec:TNG50}

TNG50 is a cosmological uniform-resolution simulation which evolves cold DM (CDM), gas, stars, super massive black holes (SMBHs), and magnetic fields in a cubic volume
of $51.7$ comoving Mpc a side from $z = 127$ to the current epoch with $\Lambda$CDM cosmology: $\Omega_m = 0.272$, $\Omega_b = 0.0455$, $\Omega_\Lambda = 0.728$, $h = 0.702$ \citep{Planck_2016}.

The astrophysical processes accounted for in TNG50 include primordial and metal-line cooling down to $10^4 \mathrm{K}$, heating from a spatially homogeneous UV/X-ray background and from localized X-ray radiation from SMBHs, density-threshold based star formation,
galactic-scale stellar-driven outflows, seeding, growth, and feedback
of SMBHs, evolution and amplification of cosmic magnetic fields via
ideal MHD, stellar evolution, stellar mass loss, and enrichment from
AGB, SNII, and SNIa tracked via nine elements (H, He, C, N, O, Ne,
Mg, Si, Fe) in addition to a subgrid model for mergers of neutrons
stars as injection sites of r-process material \citep{pillepich2023milky}. 

Importantly, in TNG50, the cold and dense phase of the ISM is not directly resolved. Instead, star-forming gas is treated with an effective model (the two-phase model of \cite{Springel_2003}, with ensuing effective equation of state for gas above $0.13 \ \mathrm{atoms}  \ \mathrm{cm}^{\text{-} 3}$), and all gas in the simulation has temperatures of $\gtrsim 10^4$ K. Stellar particles represent mono-age stellar populations with a \cite{Chabrier_2003}
initial mass function. Although the TNG50 model for stellar feedback neglects small-scale interactions due to the hydrodynamically decoupled wind particle scheme, even for SN-driven winds, the calculations do capture disordered motions indirectly induced by stellar feedback outflows \citep{TNG50_01}.

We use the MW and Andromeda analogs from the TNG50 simulation \citep{pillepich2023milky} that have been chosen according to the following criteria at $z=0$:
\begin{enumerate}
    \item Stellar mass: the galaxy stellar mass is in the following range: $\mathrm{M}_\star(< 30 \ \mathrm{kpc}) = 10^{10.5 - 11.2}\mathrm{M}_\odot$
    \item Stellar morphology: a disk-like stellar morphology needs to be
manifest, including the presence of spiral arms. This is implemented
by requiring that one or the other following criterion is fulfilled:
\begin{enumerate}
    \item the minor-to-major axis ratio of the galaxy’s stellar mass
distribution is smaller than 0.45, with the latter measured between
1 and 2 times the stellar half-mass radius: $c/a$ [stellar mass at 1-2
$r_{\star1/2}] \leq 0.45;$
\item the galaxy appears of disky shape and exhibits spiral arms by
visual inspection of three-band images of the simulated galaxies
in edge-on and face-on projections.
\end{enumerate}
\item Environment: no other galaxy with stellar mass $\geq 10^{10.5}\mathrm{M}_\odot$ is within 500 kpc distance and the total mass of the halo host is smaller than that typical of massive groups, i.e. $\mathrm{M}_{200c}(\mathrm{host}) < 10^{13} \mathrm{M}_\odot$.
\end{enumerate}

With the following constraints on the disks:
\begin{enumerate}
    \item Thin and thick disk heights consistent with those of the MW (approximately in the range $175 {-} 360$ pc and $625 {-} 1450$ pc, respectively), measured at either $7 {-} 9$ kpc.
    \item Both disk scale length and stellar mass in the ranges $ 1.7 - 2.9$ kpc and $10^{10.5-10.9}\mathrm{M}_\odot$,
\end{enumerate}
encompassing available literature constraints. 

\subsection{Auriga Simulations}

Auriga is a suite of cosmological zoom simulations of galaxy formation in isolated halos, selected from a 100 comoving Mpc box, similar in mass to that of
the MW, with the magneto-hydrodynamical moving-mesh code
AREPO \cite{Springel_2010}, which is a quasi-Lagrangian method that
solves the fluid equations on a moving mesh \citep{Auriga_2}. The halos are evolved from $z=127$ to current epoch, using the cosmological parameters: $\Omega_M = 0.307$, $\Omega_b= 0.048$, $\Omega_\Lambda = 0.693$ and a Hubble constant of $H_0 = 100 \ h \mathrm{km} \mathrm{s}^{-1} \mathrm{Mpc}^{-1}$, where $h = 0.6777$, taken from \cite{planck_2014}.

\

The Auriga physics model is broadly similar to that of IllustrisTNG \citep{TNG50_01} mentioned in section \ref{subsec:TNG50}, with four key differences \citep{Auriga_1}:
\begin{enumerate}
    \item The equation of state used for the ISM: Auriga adopts the original “stiff” form of \cite{Springel_2003}, whereas TNG uses a “softer” equation of state, which may affect the thickness of the star-forming gas layer of the disc \citep{verma_2021}.
    \item The radio mode of AGN feedback: Auriga gently adds thermal energy to random locations in the halo gas, thus inflating hot bubbles to balance X-ray losses from the halo, whereas TNG uses a kinetic jet model.
    \item Stellar wind scaling: Auriga does not impose a floor for the minimum velocity of non-local stellar winds, whereas TNG enforces a fixed wind velocity floor of $350 \mathrm{km}/\mathrm{s}$. This difference ensures slower winds for halos of mass $\lesssim 10^{11} \mathrm{M}_\odot$ in Auriga compared to TNG.
    \item Stellar yields: Auriga uses the same yield tables as the original Illustris: \cite{karkas_2010} (AGB stars); \cite{Portinari_1998} (SNII); and \cite{Thielemann:2002fw}; \cite{Travaglio_2004} (SNIa), whereas TNG uses different yield tables for SNIa and additional yield tables for some stellar mass ranges of SNII and AGB stars (see Table 2 of \citep{TNG50_01}, for more details).
\end{enumerate} 
Although each of the differences listed above has an impact on the outcome of the simulations, \citep{Auriga_1} emphasizes the similarities of the Auriga and TNG physics models. This means that the Auriga zoom-in simulations are effectively validated on large cosmological scales for a wider range of galaxy masses and types.

\begin{table*}[htbp]
  \centering
  \begin{tabular}{ccccc}
    \hline
    \textbf{Simulation} & \textbf{Halo} & \textbf{Stellar Mass ($10^{10}\mathrm{M}_\odot$)} & \textbf{${r}_{200c}$ (kpc)} & \textbf{${M}_{200c}$ ($10^{12} \mathrm{M}_\odot$)} \\
    \hline
    \textbf{FIRE 2} 
    & m12m & $11.0$ & 371 & 1.6 \\
    & m12i & $6.3 $ & 336 & 1.2 \\
    & m12b & $8.5 $ & 358 & 1.4 \\
    & m12w & $5.7 $ & 319 & 1.1 \\
    & m12f & $7.9 $ & 380 & 1.7 \\
    & m12c & $5.8 $ & 351 & 1.4 \\
    \hline
    \textbf{Auriga} 
    & 6 & $6.4   $ & 336 & 1.2 \\
    & 16 & $9.1  $ & 380 & 1.7 \\
    & 21 & $8.8  $ & 351 & 1.4 \\
    & 23 & $9.0  $ & 319 & 1.1 \\
    & 24 & $8.7  $ & 371 & 1.6 \\
    & 247 & $9.9$ & 358 & 1.4 \\
    \hline
    \textbf{{\small VINTERGATAN-GM}} 
    & 685x12 & $1.4 $ & 182.8 & 0.8 \\
    & 685x11 & $1.3 $ & 190.1 & 0.7 \\
    & 685 & $1.4$ & 191.3 & 0.7 \\
    & 685x095 & $1.2$ & 189.3 & 0.7 \\
    & 685x09 & $1.4 $ & 188.0 & 0.7 \\
    & 715 & $1.1 $ & 197.1 & 0.8 \\
    \hline
    \textbf{TNG50} 
    & 372755 & $5.5 $ & 351.0 & $4.6 $ \\
    & 502371 & $4.0$& 244.7  & $1.6$ \\
    & 535774 & $4.1 $ & 211.4 & $1.0$ \\
    & 538905 & $4.3 $ & 218.4 & $1.1$ \\
    & 550149 & $3.5 $ & 204.5 & $0.91$ \\
    & 552581 & $3.5 $ & 202.9 & $0.89$ \\
    \hline
  \end{tabular}
  \caption{List of the simulations suites along with the stellar mass enclosed within $20~\rm{kpc}$, $r_{200c}, M_{200c}$, the spherical radius and total mass at which the mean enclosed mass volume density equals 200 times the critical density for closure.}
  \label{tab:consolidated}
  
\end{table*}

\section{Adiabatic Contraction Calculation} \label{sec:AC calc}

 In the Canonical AC picture \citep{Blumenthal}, the mass distribution in the Galaxy is assumed to be spherically symmetric (where particles are on circular orbits), the DM halo is split into shells which contract homologously (i.e. mass in each shell is conserved as the halo contracts), and angular momentum is conserved. This scheme leads to $rM(r)$ as the conserved quantity throughout the evolution of the halo (see Appendix \ref{app:AC_calculation} for a derivation).

Since DM particles are on highly eccentric orbits (see \cite{eccentric_orbits}), $rM(r)$ is no longer an adiabatic invariant. The conserved quantities of eccentric orbits are angular momentum $J$ and the radial action $ I_r \equiv \frac{1}{\pi} \int_{r_p}^{r_{a}} v_r dr $. G04 argued that using the value of the mass within the average radius of a given orbit is a good proxy for the radial action. However, the orbits have a wide distribution of eccentricities which should be taken into account. This is done by averaging over the population of orbits at a given radius (see section 4.2 in G04 for more details): \begin{equation}\label{eq:rbar}
\bar{r} = \mathrm{r}_{200c} \ A x^W, \ \text{where~} x \equiv \frac{r}{\mathrm{r}_{200c}},
\end{equation}
with the conserved quantity now being $r M(\bar{r})$. The variables are defined as follows: $r_{200c}$ is the radius at which the average density equals 200 times the critical density and $(A,W)$ are halo specific dimensionless parameters.  Though the parameters $(A,W)$ differ from halo to halo, we pick an average value for each simulation suite (see Table \ref{tab:sim_parameters}). These parameters were determined by minimizing the difference between the DM density profile from the AC calculation and that from the full hydrodynamic simulation.
\begin{table}[t]
    \centering
    \begin{tabular}{ccc}
        \hline
        Simulation &  $A$ &  $W$ \\
        \hline
        TNG50 & 0.4 & 0.45 \\
        FIRE-2 & 0.36 & 0.23 \\
        {\small VINTERGATAN-GM} & 0.25 & 0.45 \\
        Auriga & 0.61 & 0.5 \\
        % Add more rows as needed
        \hline
    \end{tabular}
    \caption{Values of $A$ and $W$ used to calculate the orbit-averaged radius $\bar{r}$. }
    \label{tab:sim_parameters}
\end{table}

\subsection{Input Overview}
 The AC model posits that as the initial stellar mass profile ($M^{\text{initial}}_{\text{Stars}}$) transforms into the final stellar profile ($M^{\text{final}}_{\text{Stars}}$), the initial DM profile ($M^{\text{initial}}_{\text{DM}}$) will contract to become ($M^{\text{final}}_{\text{DM}}$) as dictated by the conserved radial action. These profiles are obtained as follows: we begin with an initial DM distribution that is obtained from the dark matter only simulation (labeled `DMO'), and assume that the initial stellar distribution is self-similar to that of the DM distribution. The stars then evolve to form the final stellar distribution, which is obtained from the full hydrodynamic simulation (labeled `hydro').
The DMO enclosed mass is normalized to match the combined DM and stellar mass from the hydro simulation at the virial radius defined as $r_{200c}$. The inputs required are summarized in Equations \ref{eq:input_1}$-$\ref{eq:input_4}.

\begin{equation} \label{eq:AC}
\begin{split}
    r_\text{initial}\big(M^{\text{initial}}_{\text{DM}}(\bar{r}_\text{initial}) 
    &+ M^{\text{initial}}_{\text{Stars}}(\bar{r}_\text{initial})\big) \\
    = r_\text{final}\big(M^{\text{final}}_{\text{DM}}(\bar{r}_\text{final}) 
    &+ M^{\text{final}}_{\text{Stars}}(\bar{r}_\text{final})\big)
\end{split}
\end{equation}
\begin{align} \label{eq:input_1}
    f_b&=\frac{\mathrm{M}_{\text{Stars}}^{\text{hydro}}(\mathrm{r}_{200c})}{\mathrm{M}_{\text{DM}}^{\text{hydro}}(\mathrm{r}_{200c})}\\
    \label{eq:input_2}
f_{\rm{norm}} &= \frac{M^{\text{hydro}}_{\text{DM}}(\mathrm{r}_{200c})+M^{\text{hydro}}_{\text{Stars}}(\mathrm{r}_{200c})}{M^{\text{DMO}}_{\text{DM}}(\mathrm{r}_{200c})}\\
\label{eq:input_3}
     M^{\text{initial}}_{\text{DM}}(r) &= (M^{\text{DMO}}_{\text{DM}}(r) \cdot f_{\text{norm}})\cdot (1-f_b)\\
     \label{eq:input_4}
      M^{\text{initial}}_{\text{Stars}}(r) &= (M^{\text{DMO}}_{\text{DM}}(r) \cdot f_{\text{norm}})\cdot f_b
\end{align}

We reiterate that $M^{\text{final}}_{\text{DM}}$ was obtained by solving Equation \ref{eq:AC} numerically using code publicly available at \href{https://github.com/abdelazizhussein/MW-Inner-DM-Profile}{\includegraphics[height=1.2em]{logos/github-mark.png}} \url{https://github.com/abdelazizhussein/MW-Inner-DM-Profile} adapted from GALPY \cite{Bovy_2015}.

\section{Flux Map and Axis Ratios}\label{sec: flux map calculation}

\subsection{Flux Map Calculation}
To calculate the $l.o.s$ integral (Equation \ref{eq:los_integral}), we place an observer in the solar circle ($8 \ \rm{kpc}$ away from the galctic center), and use \texttt{HEALPIX}\footnote{\url{http://healpix.sourceforge.net}} \cite{healpix1,healpix2} to divide the celestial sphere into $27648$ equal area pixels with angular resolution of $0.02$ rad. Below is an outline of the calculation:
\begin{itemize}
    \item Each pixel defines a $l.o.s$. Thus, we find the DM particles associated with each pixel.
    \item Next, we discretize the $l.o.s$ path into cubes of side length $\Delta x = 100 \ \rm{pc}$.
    \item Lastly, we calculate $\rho_{\rm{DM}} = \frac{\text{\# of particles in box}}{\Delta x^3}$ 
for each box and carry out the integration in Equation \ref{eq:los_integral} accordingly.

The average of $J$ factor over a solid angle  is calculated as follows:
\begin{equation} \label{eq:avg_j}
    \bar{J}(\theta_{\rm{max}}) = \frac{2\pi}{\Delta \Omega}\int_0^{\theta_{\rm{max}}} d\theta \sin\theta\int_{l.o.s} \bigg(\frac{\rho(r(s,\theta))}{\rho_\odot}\bigg) ^2 \frac{ds}{r_\odot},
\end{equation}
where $\theta$ is the opening angle of the $l.o.s$ cone, $s$ is the distance along the line of sight,  $r(s,\theta) = \sqrt{r_\odot^2+s^2-2r_\odot s \cos\theta}$ and $\Delta\Omega = 2\pi\int_0^{\theta_{\rm{max}}} d \theta \sin\theta$.

\subsection{Axis Ratios} \label{sec: axis ratios}
To quantify the morphology of the annihilation flux, ellipses are fitted to the isophotes of the surface brightness using the package \texttt{PHOTUTILS} \citep{photutils}. The results are shown in Figure \ref{fig:axis-ratios}, where the ratio of the semi-minor axis $b$ to the semi-major axis $a$, $q = b/a$, is plotted against the semi-major axis extent in angular distance for each fitted ellipse. 
\begin{figure*}[h]
    \centering
    \includegraphics[width=1\linewidth]{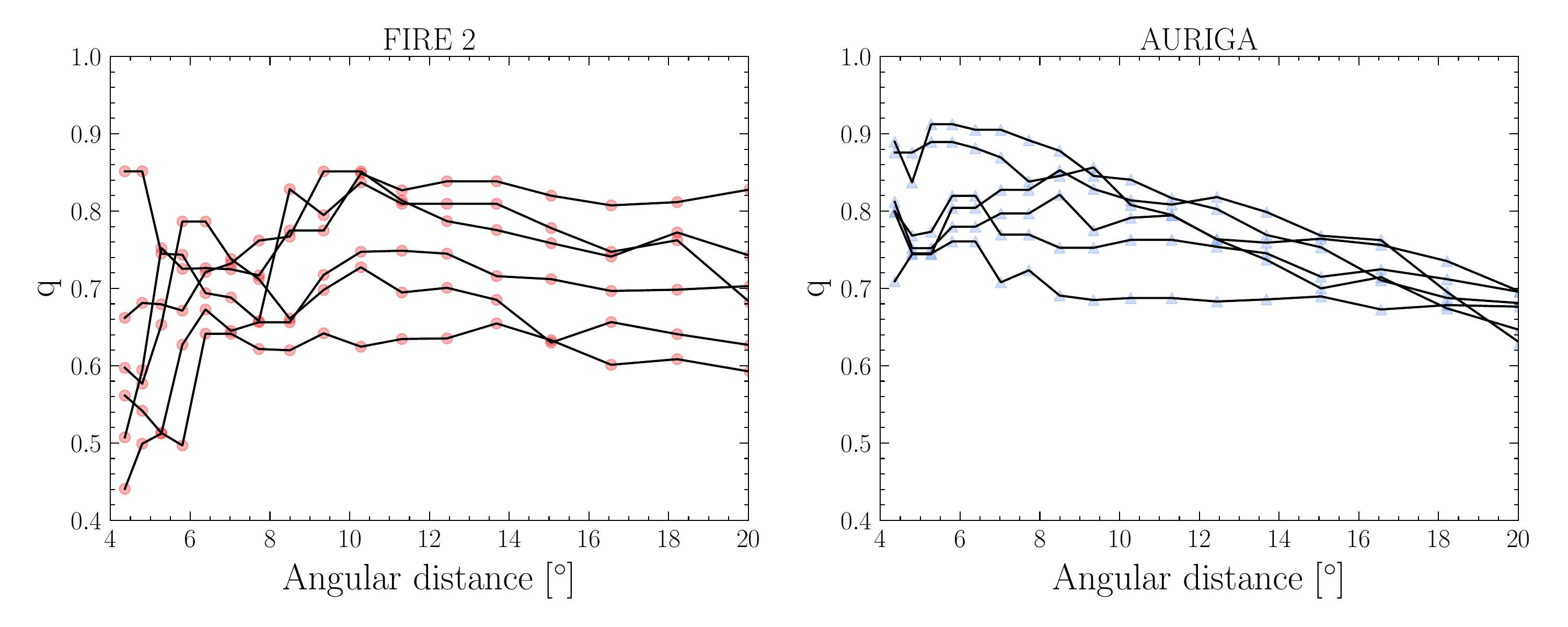}
    \caption{\label{fig:axis-ratios}Distribution of the axis ratio $q$ plotted against the size of the semi major axis in angular distance. The ellipses are fitted to the isodensity contours for all FIRE-2 and Auriga halos.}
\end{figure*}

We further explore the correlation between $\rm{q}(5^\circ)$ and the density of the stars and DM at $1 \ \rm{kpc}$ plotted in Figure \ref{fig:rho_q_correlation}. We find a weak correlation between the axis ratio and the DM density, and a stronger anti-correlation between the axis ratio and the stellar density. Though {\small VINTERGATAN-GM} was excluded from the shape analysis, we note that $\bar{\rho}_{\star}(<1 \text{ kpc}) \approx 30 \text{ GeV/}\text{cm}^3$ (averaged between the 6 simulation suites), whereas most of the simulations from Auriga and FIRE-2 have values of $\bar{\rho}_{\star}(<1 \text{ kpc})$ between 100-150$\text{ GeV/}\text{cm}^3$.

\begin{figure*}
    \centering
    \includegraphics[width=1\linewidth]{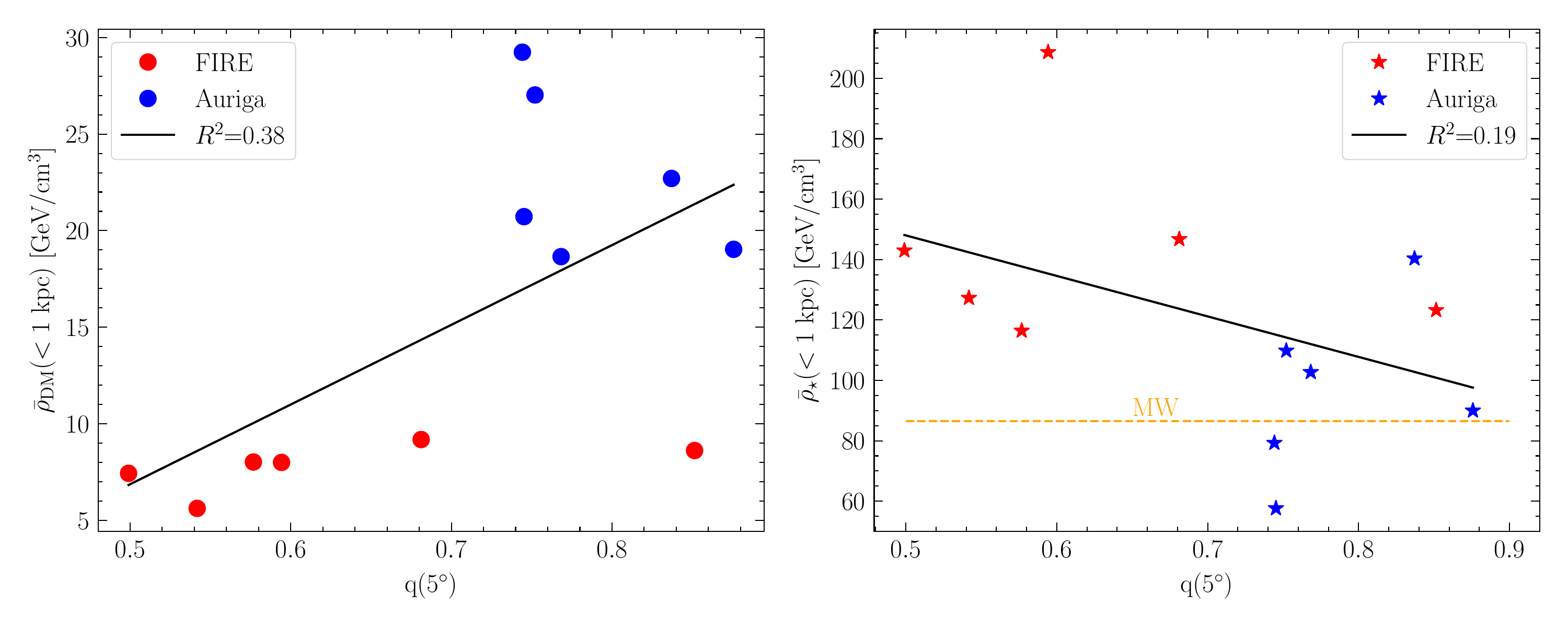}
    \caption{\label{fig:rho_q_correlation} The axis ratio at $5^\circ$ from the galactic center is plotted against $\bar{\rho}_{\rm{DM}}(<1 \ \rm{kpc})$ on the left and $\bar\rho_{\star}(<1 \ \rm{kpc})$ on the right for FIRE and Auriga. We further plot $\bar\rho_{\star}(<1 \ \rm{kpc})$ for the MW model we utilize from \cite{Model_2} using the orange dashed line. We calculate the coefficient of determination as a goodness of fit parameter and present it in the legend. A weak correlation of between the axis ratio and the DM density is observed while we note a stronger anti-correlation with the stellar density.} 
\end{figure*}

\end{itemize}

\subsection{{\small VINTERGATAN-GM} Flux} \label{sec:Vintergatan_flux}
\begin{figure}[h]
    \centering
    \includegraphics[width=0.9\linewidth]{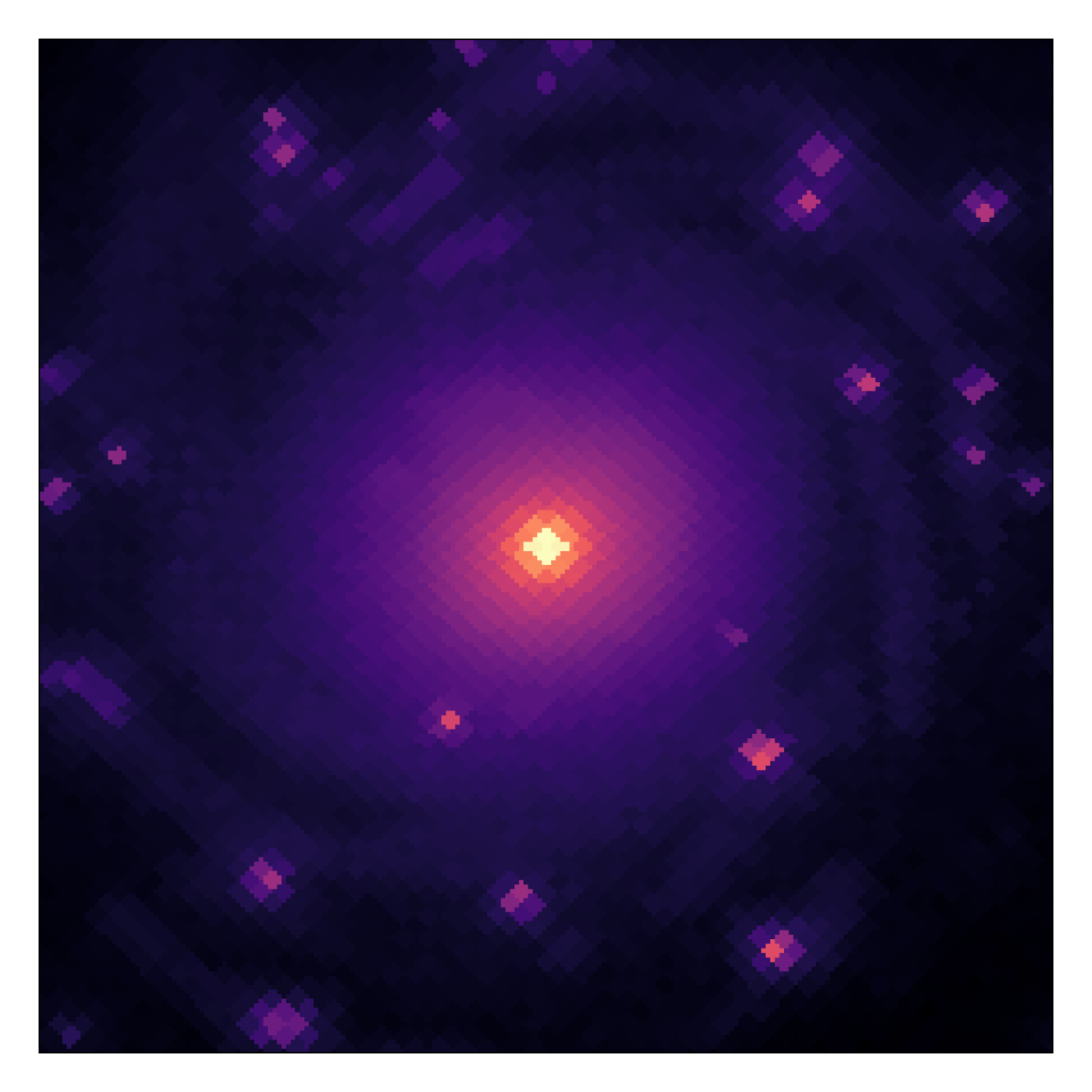}
    \caption{$40\times 40^\circ$ Annihilation flux plot for Halo 685. The presence of many subhalos prevents us from extracting shape information from fitting isodensity contours.}
    \label{fig:vintergatan_sub}
\end{figure}

We calculate the annihilation flux map for the {\small VINTERGATAN-GM} simulation suite and plot Halo 685 as an example in Figure \ref{fig:vintergatan_sub}. We conclude that the presence of a large number of subhalos near the galactic center prevents us from making conclusions about the shape of the halos by fitting ellipses to isodensity contours. More explicitly, we find 3 halos without subhalos within 15 degrees of the GC, 2 of them exhibit a spherical morphology while the other exhibits a more oblate morphology.

%TC:endignore

\end{document}